%% file: main.tex
\documentclass[preprint, aps, pra, showpacs, floatfix, superscriptaddress]{revtex4-1}
\usepackage{graphicx}
\usepackage{amsthm,amsfonts,amssymb,verbatim}
\usepackage[usenames]{color}    
\usepackage{colortbl}
\usepackage{calrsfs}
\usepackage{hyperref} 
\usepackage{subfigure}
\usepackage{indentfirst}
\usepackage{pgfplots}

\usepackage{mathtools}
\usepackage{physics}
\usepackage{xcolor}
\usepackage{adjustbox}
\usepackage{placeins}
\usepackage[T1]{fontenc}
\usepackage{lipsum}
\usepackage{csquotes}
\usepackage{fancyhdr}

\usepackage{cmap} 				
\usepackage{siunitx} 

\newcommand{\bp}{{\bf p}}
\newcommand{\bk}{{\bf k}}
\newcommand{\br}{{\bf r}}
\newcommand{\hbr}{{\hat{\bf r}}}
\newcommand{\hbp}{{\hat{\bf p}}}
\newcommand{\ba}{{\bf a}}
\newcommand{\bY}{{\bf Y}}
\newcommand{\balpha}{\boldsymbol{\alpha}}
\newcommand{\bsigma}{\boldsymbol{\sigma}}
\newcommand{\bepsilon}{\boldsymbol{\epsilon}}
\newcommand{\bphi}{\boldsymbol{\phi}}
\begin{document}
\title{
Two-photon Annihilation of Positrons with K-shell Electrons of H-like ions
}

\author{Z.~A.~Mandrykina}
\affiliation{
Department of Physics, St. Petersburg State University,
Universitetskaya naberezhnaya 7/9, 199034 St. Petersburg, Russia
}
\author{V.~A.~Zaytsev}
\affiliation{
Department of Physics, St. Petersburg State University,
Universitetskaya naberezhnaya 7/9, 199034 St. Petersburg, Russia
}
\author{V.~A.~Yerokhin}
\affiliation{
Peter the Great St.~Petersburg Polytechnic University,
St. Petersburg 195251, Russia
}
\author{V.~M.~Shabaev}
\affiliation{
Department of Physics, St. Petersburg State University,
Universitetskaya naberezhnaya 7/9, 199034 St. Petersburg, Russia
}

\date{\today} 

\begin{abstract}
The two-photon annihilation of a positron with an electron bound in the $1s$ state of a H-like ion is calculated within the fully relativistic QED framework.
The interaction with the nucleus is treated nonperturbatively, thus allowing the calculations to be carried out for the annihilation with strongly-bound inner shells of heavy ions.
Infrared divergences, appearing when one of the emitted photons approaches the low-frequency limit, are accurately eliminated from final expressions.
The total cross section of the two-photon and one-photon annihilation processes are compared for a wide range of collision energies and nuclear charge numbers.
It is demonstrated that the two-photon annihilation channel dominates over the one-photon channel for the low and medium-$Z$ ions, whereas for
the high-$Z$ ions the situation reverses. 
\end{abstract}


\maketitle

\input{sections/introduction}
\input{sections/theory}
\input{sections/num_det}
\input{sections/res_disc}
\input{sections/summary}
\input{sections/acknowledgements.tex}


\end{document}

%% file: sections/introduction.tex
\section{Introduction}
\label{sec1}
The electron-positron annihilation is one of the fundamental processes of matter-antimatter interaction. It attracted interest of investigators for a very long time and yielded a number of important fundamental results. Among them was one of the first demonstrations of the violation of Bell's inequalities
from studies of polarization correlations between the high-energy photons produced during the positronium annihilation \cite{KUW, WLB, BAM}.
Investigations of the electron-positron annihilation have also many practical applications. This process was used as a valuable tool for studying defects in metals and semiconductors~\cite{Weiss,Tuomisto_RMP85_1583_2013}, performing positron-emission tomography~\cite{PET_2005,PET_2006}, facilitating astrophysical searches~\cite{Guessoum_AA436_171_2005, Lingenfelter_PRL103_031301_2009, Prantzos_RMP83_1001_2011}, and other applications \cite{Surko_JPB38_R57_2005, Weiss_RPC76_285_2007, Hugenschmidt,Cizek_NJP14_035005_2012}.
An important scenario is annihilation of positrons on the inner-shell electrons of an atomic or ionic target. This process allows one to study the matter-antimatter interaction in the presence of the strong Coulomb field of the nucleus. 
Although a direct measurement of the annihilation with selected inner shells is a difficult task, such experiments were successfully held in the past~\cite{Nagatomo_1974, PRL_Hunt, PRL_Eshed, Kim}.
A new generation of such experiments is going to become possible in the near future, specifically, at the Lawrence Livermore National Laboratory, the ELI-NP Research Center, and the future FAIR facility.
\\
\indent
The annihilation of a positron and a bound electron can proceed with the emission of one, two, or more photons. It is well known that in the absence of the nucleus, the single-quantum annihilation is forbidden because of the energy-momentum conservation requirements. Therefore, one can expect this channel to be strongly suppressed for light atoms, where the Coulomb field by the nucleus only weakly violates the free-space momentum conservation law. The two-photon annihilation is allowed and typically dominates over other channels in an empty space. Its cross section, however, is suppressed by an additional power of the fine-structure constant as compared to the single-quantum cross section. With this in mind, one can expect that the two-photon annihilation dominates over the one-photon channel for light systems, whereas for heavy systems the situation reverses~\cite{Drukarev_2016}. This assumption, however, has never been confirmed by accurate calculations or experimental investigations.
\\
\indent
In order to provide reliable theoretical predictions for the annihilation cross section with high-$Z$ ions, one needs to perform calculations within a fully relativistic QED formalism and to all orders in the binding field of the nucleus. 
\\
\indent
For the one-photon annihilation, such treatment was first developed by Johnson and co-authors~\cite{Johnson_PR135_A1232_1964, Johnson_PR159_61_1964}. At present, such calculations are well established. For the two-photon annihilation with bound electrons, previous studies were performed for the two extreme cases of the ultraslow  \cite{Chang_ZETF33_365_1957} and ultrafast \cite{Dirac,Tamm} positrons. To the best of our knowledge, the only rigorous QED calculation of this process was performed recently by some of us in Ref.~\cite{Zaytsev_PRL123_093401_2019}. In that work, the finite-basis-set approach was used for the construction of the virtual electron-positron state propagator. The applicability of this approach turned out to be limited by the restriction for the energy of the electron-positron propagator to be above the negative-energy continuum threshold. As a result, not all possible combinations of energy sharing between the two emitted photons can be described and no reliable data can be obtained for positrons with energies larger than a few hundred keV. Calculations for these energies are, however, required for the comparison of single- and double-quanta annihilation in high-$Z$ ions.
\\
\indent
In the present investigation we aim to overcome the limitations of the approach of Ref.~\cite{Zaytsev_PRL123_093401_2019}. This is achieved by representing the electron-positron propagator in the presence of the binding nuclear field with the exact Dirac-Coulomb Green function. It should be noted that this extension of the method is associated with significant technical difficulties. One of the reasons is that the electron-positron propagator for energies beyond the continuum threshold is a strongly oscillating and slowly decreasing function for large radial distances. This calls for special numerical techniques for computation of radial integrals.  Furthermore, special care needs to be taken in the region where one of the emitted photons approaches the low-frequency limit, because of infrared divergences. In the present work we overcome all the difficulties and evaluate the total cross section for the two-photon annihilation of positrons with $1s$ electrons in a wide range of positron energies and nuclear charge numbers $Z$.
\\
\indent
The outline of the paper is as follows.
In Sec.~\ref{sec2A} we recall basic relations for the one-photon annihilation process.
Section~\ref{sec2B} represents the theoretical description of the two-photon channel.
In Sec.~\ref{sec2C} we discuss in detail the infrared divergences arising in the double-quanta annihilation.
Sec.~\ref{sec3} presents numerical details of the calculation.
The total cross section for the one- and two-photon annihilation of the positrons with the $1s$ electrons of the H-like ions are presented in Sec.~\ref{sec4}.
Section~\ref{sec5} summarizes and concludes the paper.
\\
\indent
Relativistic units $(m_e = \hbar = c = 1)$ and the Heaviside charge units $(e^2 = 4\pi\alpha)$ are utilized throughout the paper. 

%% file: sections/theory.tex
\section{BASIC FORMALISM}
\label{sec2}
%
%
In this section we present basic formulas for the one- and two-photon annihilation of a positron with an electron bound in an atom within the framework of the rigorous QED approach. 
%
%
%
\subsection{One-photon annihilation}
\label{sec2A}
%
%
The amplitude for the one-photon annihilation of a positron with an electron in a bound $a$ state is given by
\begin{equation}
\label{tau}
\tau^{(\rm{1 ph})}_{\lambda, \mu_i m_a}
=
\left\langle (-p_i\mu_i) \left\vert
\balpha \cdot {\bf A}^{*}_{\bk\lambda}
\right\vert  a m_a \right\rangle,
\end{equation}
where $p_i$ and $\mu_i$ are the asymptotic four-momentum and helicity of the incoming positron, respectively, and $m_a$ is the total angular-momentum projection of the bound electron.
The wave function of the plane-wave photon with the energy $\omega$, the momentum $\bk$, and the polarization $\lambda$ is given by
\begin{equation}
\label{eq_A}
{\bf A}_{\bk\lambda}
\equiv
{\bf A}_{\bk\lambda} (\br)
=
\frac{\bepsilon_{\bk\lambda} e^{i \bk \cdot \br}}{\sqrt{2\omega(2\pi)^3}},
\end{equation}
with $\balpha$ standing for the vector incorporating the Dirac matrices and $\bepsilon_{\bk\lambda}$ designating the circular polarization vector in the Coulomb gauge.
For practical purposes, it is convenient to represent the photon wave function as the multipole series~\cite{Rose_RET}
\begin{equation}
\label{eq_multipole_expansion}
\bepsilon_\lambda e^{i\bk\cdot\br}
=
\sqrt{2\pi}
\sum\limits_{LM_L}
i^L
\sqrt{2L+1}
D^{L}_{M_L\lambda}(\varphi_{k},\theta_{k},0)
\sum\limits_{p=0,1}
(i\lambda)^p \ba_{LM_L}^{(p)}(\br),
\end{equation}
where $\ba_{LM_L}^{(p)}$ are the magnetic ($p=0$) and electric ($p=1$) vectors
\begin{equation}
\begin{aligned}
\ba_{LM_L}^{(0)}(\br)
&=
j_L(\omega r)\bY_{LLM_L}(\hat\br),
\\
\ba_{LM_L}^{(1)}(\br)
&=
\sqrt{\frac{L+1}{2L+1}}
j_{L-1}(\omega r)\bY_{LL-1M_L}(\hat\br)
-
\sqrt{\frac{L}{2L+1}}
j_{L+1}(\omega r)\bY_{LL+1M_L}(\hat\br)
\end{aligned}
\label{eq:magn_elec_vec}
\end{equation}
with $j_L$ standing for the spherical Bessel function of the first kind~\cite{Abramovitz}
and $\bY_{JLM}$ being the vector spherical harmonics~\cite{Varshalovich_QTAM_1988}.
\\
\indent
To describe the incoming positron with the asymptotic four-momentum $p_i$ and the helicity $\mu_i$ it is more convenient to represent it as an outgoing electron with the asymptotic four-momentum $-p_i$ and the same helicity $\mu_i$~\cite{Bjorken, Itzykson}.
The wave function of such electron is given by~\cite{Rose_RET,Artemyev_PRA79_032713_2009}
\begin{equation}
\label{phi1}
\Psi^{(-)}_{-p_i \mu_i}(\mathbf{r})
=
\frac{1}{\sqrt{4\pi \varepsilon_i \vert\bp_i\vert}}
\sum_{\kappa m_j}
i^{-\bar{l}}
e^{-i\delta^{(V)}_{-\varepsilon_i\kappa}}
\sqrt{2\bar{l} + 1}
C^{j-\mu_i}_{\Bar{l} 0\, 1/2 -\mu_i}
D^{j}_{m_j -\mu_i}(\varphi_{\hat{\bp}_i}, \theta_{\hat{\bp}_i}, 0)
\Psi^{(V)}_{-\varepsilon_i \kappa m_j}(\br),
\end{equation}
where $\bp_i$ and $\varepsilon_i = \sqrt{\bp_i^2 + 1}$ are the asymptotic momentum and energy of the positron, respectively, $\kappa = (-1)^{l+1/2-j}(j+1/2)$ is the Dirac quantum number determined by the total $j$ and orbital $l$ angular momenta, $m_j$ is the projection of the total angular momentum, $\bar{l} = 2j - l$, $\delta^{(V)}_{-\varepsilon_i \kappa}$ is the phase shift being induced by central potential $V$, $C^{JM}_{j_1m_1\, j_2m_2}$ is the Clebsch-Gordan coefficient, $D^{J}_{MM'}$ is the Winger matrix \cite{Rose_ETAM_1957,Varshalovich_QTAM_1988}, and $\varphi_{\hat{\bp}_i}$ and $\theta_{\hat{\bp}_i}$ denote the azimuthal and polar angles of the unit vector $\hat{\bp}_i = \bp_i / \vert\bp_i\vert$.
The partial waves
\begin{equation}
\Psi^{(V)}_{-\varepsilon_i\kappa m_j}(\br) =
\begin{pmatrix}
g^{(V)}_{-\varepsilon_i\kappa}(r) \Omega_{\kappa m_j}(\hat{\br})
\\
if^{(V)}_{-\varepsilon_i\kappa}(r) \Omega_{-\kappa m_j}(\hat{\br})
\end{pmatrix}
\end{equation}
are the negative-energy continuum solutions of the Dirac equation in the central potential $V$, with $g^{(V)}_{-\varepsilon_i\kappa}$ and $f^{(V)}_{-\varepsilon_i\kappa}$ standing for the large and small radial components, whose explicit form for the pure Coulomb potential can be found, e.g., in Refs~\cite{Akhiezer_1965,Berestetsky_2006}, $\Omega_{\kappa m_j}$ is the spherical spinor~\cite{Varshalovich_QTAM_1988}, and $\hat{\br}$ is the unit vector into the $\br$ direction.
\\
\indent
Substituting Eqs.~\eqref{eq_A} and~\eqref{phi1} into Eq.~\eqref{tau} and utilizing the multipole expansion~\eqref{eq_multipole_expansion}, we obtain the expression for the amplitude in a form appropriate for the direct numerical calculations.
The total cross section (TCS) is connected to the amplitude as follows
\begin{equation}
\label{tot}
\sigma^{(\rm{1 ph})}_{\rm{tot}}
=
2\alpha \omega^2 \frac{(2 \pi)^5}{v_i}
\frac{1}{2(2j_a + 1)} \sum_{\mu_i m_a} \sum_{\lambda} \int d\Omega_k
\left\vert
\tau^{(\rm{1 ph})}_{\lambda, \mu_i m_a}
\right\vert^2,
\end{equation}
where $v_i$ is the velocity of the incoming positron and $j_a$ is the total angular-momentum of the bound electron.
%
%
\subsection{Two-photon annihilation}
\label{sec2B}
%
%
%
The amplitude for the positron--bound-electron annihilation with the emission of two photons enumerated by subscripts $1$ and $2$ is defined by the diagrams shown in Fig.~\ref{2ph}, which lead to the following expression~\cite{Akhiezer_1965, Berestetsky_2006}:
\begin{figure}[h!]
\includegraphics[width=0.48\textwidth]{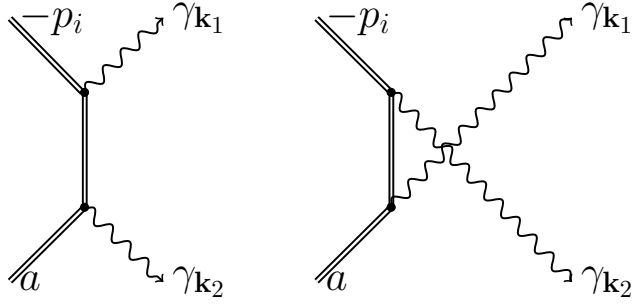}
\caption{
Feynman diagrams for the two-photon annihilation of the positron $e^{+}_{\mathbf{p}_i}$ with the bound electron in the $a$ state.
The double lines indicate the positron, virtual, and electron states in the field of the nucleus.
The wavy lines represent the emitted photons, $\gamma_{\bk_1}$ and $\gamma_{\bk_2}$.
}
\label{2ph}
\end{figure}
\begin{eqnarray}
\label{tau2ph}
\nonumber
\tau^{\rm{(2 ph)}}_{\lambda_1\lambda_2,\mu_i m_a}
& = &
-\left\langle (-p_i\mu_i) \left\vert
\left( \balpha \cdot {\bf A}^*_{\bk_1\lambda_1} \right)
G(E_a - \omega_2)
\left( \balpha \cdot {\bf A}^*_{\bk_2\lambda_2} \right)
\right\vert a m_a \right\rangle
\\ &&
-\left\langle (-p_i\mu_i) \left\vert
\left( \balpha \cdot {\bf A}^*_{\bk_2\lambda_2} \right)
G(E_a - \omega_1)
\left( \balpha \cdot {\bf A}^*_{\bk_1\lambda_1} \right)
\right\vert a m_a \right\rangle,
\end{eqnarray}
where $E_a$ is the energy of the bound electron.
The Dirac-Coulomb Green function $G(E)$ is given by~\cite{Mohr_PR293_227_1998}
\begin{eqnarray}
\label{eq_G}
\nonumber
G(E)
& \equiv &
G(E,\br_1,\br_2)
\\
& = &
\sum_{\kappa m_j}
\begin{pmatrix}
  G^{11}_\kappa(E,r_1,r_2) \Omega_{\kappa m_j}(\hbr_1)\Omega^\dagger_{\kappa m_j}(\hbr_2) &
-iG^{12}_\kappa(E,r_1,r_2) \Omega_{\kappa m_j}(\hbr_1)\Omega^\dagger_{-\kappa m_j}(\hbr_2)
\\
iG^{21}_\kappa(E,r_1,r_2) \Omega_{-\kappa m_j}(\hbr_1)\Omega^\dagger_{\kappa m_j}(\hbr_2) &
 G^{22}_\kappa(E,r_1,r_2) \Omega_{-\kappa m_j}(\hbr_1)\Omega^\dagger_{-\kappa m_j}(\hbr_2)
\end{pmatrix}.
\end{eqnarray}
Here
\begin{eqnarray}
\nonumber
G^{ij}_\kappa(E,r_1,r_2)
=
-\frac{1}{\Delta_\kappa(E)}
&&
\left[
\phi^{\infty,i}_\kappa(E,r_1) \phi^{0,j}_\kappa(E,r_2)\theta\left(r_1 - r_2\right)
\right.
\\
 + && \left.
\phi^{0,i}_\kappa(E,r_1) \phi^{\infty,j}_\kappa(E,r_2)\theta\left(r_2 - r_1\right)
\right]
\end{eqnarray}
is the radial Dirac-Coulomb Green function expressed in terms of the two-component solutions of the radial Dirac equation regular at the origin,
\begin{equation}
\label{eq_reg_0}
\bphi^0_\kappa = \begin{pmatrix}\phi^{0,1}_\kappa(E,r)\\\phi^{0,2}_\kappa(E,r)\end{pmatrix},
\end{equation}
and at infinity,
\begin{equation}
\label{eq_reg_8}
\bphi^\infty_\kappa = \begin{pmatrix}\phi^{\infty,1}_\kappa(E,r)\\\phi^{\infty,2}_\kappa(E,r)\end{pmatrix}.
\end{equation}
The Wronskian of these solutions is given by
\begin{equation}
\Delta_\kappa(E)
=
r^2
\bphi^{0^T}_\kappa(E,r)
\begin{pmatrix} 0 & -1 \\ 1 & 0 \end{pmatrix}
\bphi^{\infty}_\kappa(E,r).
\end{equation}
The explicit form of the solutions~\eqref{eq_reg_0} and~\eqref{eq_reg_8} can be found in Refs.~\cite{Mohr_PR293_227_1998,Yerokhin_PRA60_800_1999}.
\\
\indent
The single differential cross section is obtained from the amplitude (\ref{tau2ph}) as
\begin{equation}\label{DCSeq}
\frac{d\sigma^{\rm{(2ph)}}}{d\omega_1}
=
4 \alpha^2 \frac{(2\pi)^6}{v_i}
\omega^2_1 \omega^2_2
\frac{1}{2(2j_a+1)}
\sum_{\lambda_1 \lambda_2}
\sum_{\mu_i m_a}
\int d\Omega_1 d\Omega_2
\left\vert
\tau^{\rm{(2 ph)}}_{\lambda_1\lambda_2,\mu_i m_a}
\right\vert^2.
\end{equation}
The above expression for the differential cross section is infrared (IR) divergent at the endpoints $\omega_1 \sim 0$ and $\omega_1 \sim E_{\rm tot} = E_a + \varepsilon_i$, which correspond to the two cases when one of the photons carries away almost the whole energy. These divergences need to be separated out before the numerical evaluation is performed.
%
%
\subsection{Infrared divergences}
\label{sec2C}
\begin{figure}[h!]
\includegraphics[width=0.48\textwidth]{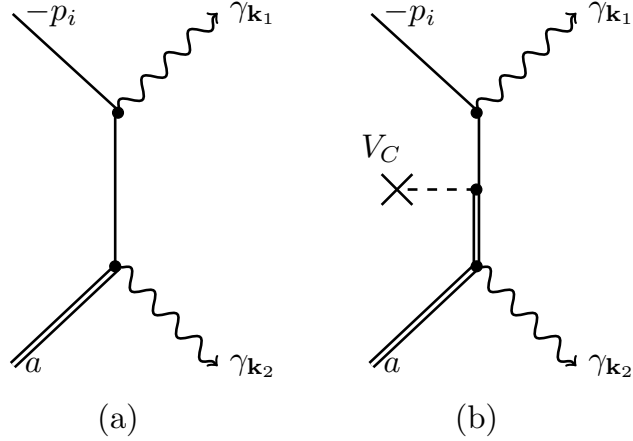}
\caption{
The infrared divergent diagrams of the two-photon annihilation at $\omega_1 \rightarrow 0$.
The single line denotes a free electron and the dashed line ended by the cross denotes the interaction with the Coulomb field.
}
\label{2ph_ir}
\end{figure}

In the present investigation, the electron-positron annihilation is described by using the perturbation expansion in powers of $\alpha$, which leads to the
series with individual terms corresponding to the emission of one, two or more photons.
The perturbation expansion is applicable when the probability of the multiple quanta emission decreases with the increase of their number.
However, this is not the case for the processes involving soft photons.
Indeed, the number of quanta carrying away the energy $\omega$ tends to infinity when $\omega\rightarrow 0$~\cite{Bloch_1937, Akhiezer_1965, Berestetsky_2006}.
A manifestation of the nonperturbative regime is the fact that the individual terms of the perturbative expansion become infrared divergent at $\omega\rightarrow 0$.
\\
\indent
Naturally, the sum of all perturbation series should be infrared finite. Moreover, it can be shown  \cite{Jauch_1976, Yennie_AP13_379_1961} that in each order of $\alpha$ the IR divergent contributions to the cross section related to the soft real and virtual photons eliminate each other. Specifically,
in the case under consideration, the IR divergent part of the two-photon annihilation should be cancelled by the corresponding contribution from the first-order radiative correction to the one-photon annihilation. The analogous cancellation of the IR divergences was demonstrated in Ref.~\cite{Shabaev_PRA61_052112_2000} in the context of the QED corrections to the radiative recombination. Calculation of QED corrections to the single-quantum annihilation lies beyond the scope of the present investigation. Therefore, we regularize the obtained formula for the two-photon annihilation amplitude by separating out the IR divergent contribution.
\\
\indent
We now obtain an explicit expression for the IR divergent contribution appearing in the differential cross section at $\omega_1 \rightarrow 0$.
For this purpose, we decompose the diagrams shown in Fig.~\ref{2ph} in powers of the interaction with the Coulomb potential and retain only the IR divergent terms. These terms are depicted in Fig.~\ref{2ph_ir} and correspond to the following expression
\begin{eqnarray}
\nonumber
\label{2ph_t}
\tau^{(\rm 2 ph,\ IR)}_{\lambda_1, \lambda_2, \mu_i, m_a} 
& = &
\tau^{(\rm{a})} + \tau^{(\rm{b})}
\\
& = &
- \left\langle V_{-p_i\mu_i}
\left\vert
\left(\balpha \cdot {\bf A}^*_{\bk_1 \lambda_1}\right)
G_0( - \varepsilon_i + \omega_1)
\left[1 + V_C G(-\varepsilon_i + \omega_1)\right]
\left(\balpha \cdot {\bf A}^*_{\bk_2 \lambda_2}\right)
\right\vert a m_a \right\rangle,
\end{eqnarray}
with the free electron-positron Green function
\begin{equation}
\label{fp}
G_0(E) = 
\int d\bp
\sum_\mu
\left[
\frac{U_{p\mu} U^\dagger_{p\mu}}{ E - p^0(1 - i0)}
+ 
\frac{V_{-p\mu} V^\dagger_{-p\mu}}{ E + p^0(1 - i0)}\right].
\end{equation}
Here $p^0 = \sqrt{\bp^2 + 1}$,
\begin{equation}
U_{p\mu} \equiv
U_{p\mu} (\br) = 
\frac{e^{i\bp\cdot\br}}{\sqrt{2p^0 (2\pi)^3}}
\begin{pmatrix} 
\sqrt{p^0 + 1} \chi_{1/2 \mu} (\hbp) 
\\ 
\sqrt{p^0 - 1} (\hbp \cdot \bsigma) \chi_{1/2 \mu}(\hbp)
\end{pmatrix},
\end{equation}
\begin{equation}
V_{-p\mu} \equiv
V_{-p\mu} (\br) = 
\frac{e^{- i\bp\cdot\br}}{\sqrt{2p^0(2\pi)^3} }
\begin{pmatrix} 
\sqrt{p^0 - 1} (\hbp \cdot \bsigma)\chi_{1/2 -\mu} (\hbp) 
\\  
\sqrt{p^0 + 1}\chi_{1/2 -\mu}(\hbp)
\end{pmatrix}
\end{equation}
are the wave functions of the free electron and free positron, respectively, and $\chi_{1/2\mu}(\hbp)$ stands for the eigenfunction of the helicity operator $(\hbp \cdot \bsigma)/2$ with the eigenvalue $\mu$. 
To extract the dominant contribution in the limit $\omega_1 \rightarrow 0$, it is sufficient to keep only the second term in Eq.~\eqref{fp} and to neglect $\omega_1$ in $G$.
Making use of these assumptions and the relation
\begin{equation}
\Psi^{(-)\dagger}_{-p_i \mu_i} = 
V^\dagger_{-p_i\mu_i} + 
V^\dagger_{-p_i\mu_i} V_C G(-\varepsilon_i),
\end{equation}
we arrive at
\begin{equation}
\label{tauIR}
\tau^{(\rm 2 ph,\ IR)}_{\lambda_1\lambda_2,\mu_i m_a}
=
\frac{1}{\varepsilon_i} \frac{\bp_i \cdot \bepsilon^*_{\bk_1\lambda_1}}{\sqrt{2\omega_1(2\pi)^3}}
\frac{1}{-\varepsilon_i + \omega_1 + \sqrt{(-\bp_i + \bk_1)^2 + 1}}
\tau^{(\rm{1 ph})}_{\lambda_2, \mu_i m_a}.
\end{equation}
The related contribution to the cross section is given by
\begin{eqnarray}
\label{ir_contrib}
\nonumber
\frac{d\sigma^{(\rm 2 ph,\ IR)}}{d\omega_1}
& = &
\sigma^{\rm(1 ph)}_{\rm{tot}}\frac{4\pi\alpha}{\varepsilon_i^2}\int d\Omega_1\sum_{\lambda_1}|\boldsymbol{\rm{p}}_i\cdot\boldsymbol{\epsilon}^{*}_{\boldsymbol{\rm{k}}_1\lambda_1}|^2\frac{\omega_1}{2(2\pi)^3(-\varepsilon_i+\omega_1+\sqrt{ (-\boldsymbol{\rm{p}}_i+\boldsymbol{\rm{k}}_1)^2+1}\big)^2}
\\
& = &
\sigma^{\rm(1 ph)}_{\rm{tot}}
\frac{I_{p_i}}{\omega_1},
\end{eqnarray}
where
\begin{equation}
\label{intensity}
I_{p_i} =
\frac{\alpha}{|\boldsymbol{\rm{p}}_i|\pi}
\left[
\varepsilon_i\ln{\Bigg(\frac{\varepsilon_i+|\boldsymbol{\rm{p}}_i|}{{\varepsilon_i-|\boldsymbol{\rm{p}}_i|}}\Bigg)}-2|\boldsymbol{\rm{p}}_i| \right].
\end{equation}
We now define the regularized two-photon annihilation differential cross-section as follows
\begin{equation}
\label{dcross_IR_free}
\frac{d\tilde{\sigma}^{\rm (2ph)}}{d\omega_1}
=
\frac{d\sigma^{\rm (2ph)}}{d\omega_1} - \sigma^{\rm(1 ph)}_{\rm{tot}} I_{p_i}
\left(
\frac{1}{\omega_1} + \frac{1}{E_{\rm tot} - \omega_1}
\right).
\end{equation}
Here the first and second terms in the brackets remove the IR divergences at the endpoints $\omega_1 \sim 0$ and $\omega_1 \sim E_{\rm tot}$, respectively.
The corresponding total cross section is given by
\begin{equation}
\label{cross_2ph}
\sigma^{\rm (2ph)}_{\rm tot} =
\frac{1}{2}
\int^{E_{\rm tot}}_0 d\omega_1
\frac{d\tilde{\sigma}^{\rm (2ph)}}{d\omega_1},
\end{equation}
where $\frac{1}{2}$ factor in front of the integral is introduced to account for the indistinguishability of the photons~\cite{Akhiezer_1965}.
The regularized differential cross section~\eqref{dcross_IR_free} tends to zero at the endpoints, which is consistent with the general expectations. 
Indeed, the annihilation probability has to be finite regardless of the number of quanta to be emitted, and since the number of quanta carrying away the energy $\omega\rightarrow 0$ tends to infinity, the probability of the emission of a single photon with $\omega\rightarrow0$ tends to zero. 

%% file: sections/num_det.tex
\section{Numerical evaluation} \label{sec3}
The numerical evaluation of the annihilation amplitude \eqref{tau2ph} is based on the computation of the Dirac-Coulomb Green function $G(E,r_1,r_2)$.
For energies below the continuum threshold, $|E| < 1$, the Dirac-Coulomb Green function can be conveniently computed either by a finite-basis-set representation or by the exact representation in terms regular and irregular Dirac-Coulomb functions, see, e.g., a review \cite{Yerokhin_20}. The comparison of the results obtained with two different approaches in this region was used as an important cross-check of the numerical procedure. For $|E| > 1$, however, the finite-basis-set representation is not applicable for the process under consideration, so one has to compute the Dirac-Coulomb functions.
\\
\indent
The most problematic region for the numerical computation is $|E| \gtrsim1$. When the Dirac-Coulomb functions are represented in terms of
the Whittaker functions $M_{\alpha,\beta}$ and $W_{\alpha,\beta}$, the region $|E| \gtrsim1$ corresponds to large and complex values of the first index $\alpha$. In this region most of the standard numerical algorithms used in the literature for the computation of the Whittaker functions \cite{Mohr_PR293_227_1998,Yerokhin_PRA60_800_1999} are not good enough. For this reason, in the present work we computed the regular and irregular solutions of the Dirac equation by numerically solving the differential equation on a radial grid, with the method described in Appendix of Ref.~\cite{Yerokhin_11_FNS}.
\\
\indent
Substituting Eqs.~\eqref{eq_A},~\eqref{phi1}, and~\eqref{eq_G} into Eq.~\eqref{tau2ph} and utilizing the multipole expansion~\eqref{eq_multipole_expansion}, the amplitude is represented as an infinite series over the multipole components of the positron ($\kappa$), the Green function ($\kappa_g$), and photons ($L_1$ and $L_2$). The summations over the photon multipoles $L_1$ and $L_2$ are finite after the angular-momentum selection rules are taken into account. The summations over $\kappa$ and $\kappa_g$, however, are infinite and need to be truncated. In our calculations, we typically truncated the expansion at $|\kappa| = 40$ and $|\kappa_g| = 45$.
\\
\indent
In each term of the partial-wave expansion, the angular integration can be separated out and calculated analytically by the standard Racah-algebra technique.
The integration over the radial variables has to be carried out numerically.
This numerical integration is quite straightforward for the energy argument of the Green function $|E| < 1$.
For $|E| > 1$, however, the integrand becomes a strongly oscillating and slowly decreasing function for large radial distances, so that the straightforward numerical integration methods fail to converge. Specifically, we need to integrate the product of the continuum-state Dirac wave function, the Dirac-Coulomb Green function with the energy $|E| > 1$, and the spherical Bessel function. To this end, we use the method of the rotation of the integration contour in the complex $r$ plane, which transforms the integrand to a smooth and exponentially decaying function at large $r$. This method was developed in Ref.~\cite{Yerokhin_10_BS} in the context of bremsstrahlung and later extended in Ref.~\cite{Yerokhin_PRA84_032703_2011} for the double photoionization process.

%% file: sections/res_disc.tex
\section{RESULTS AND DISCUSSION} 
\label{sec4}
\begin{figure}[h!]
\includegraphics[width=0.8\linewidth]{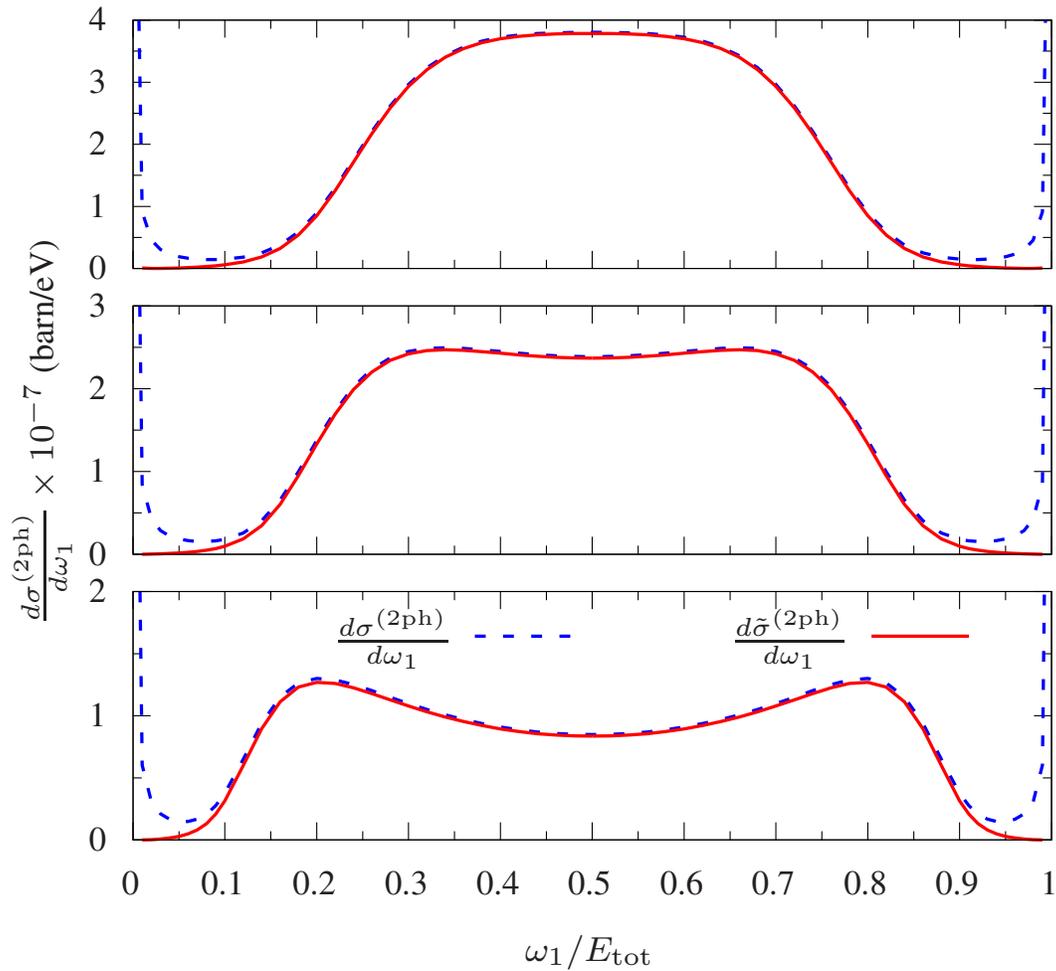}
\caption{
Differential cross section for the two-photon annihilation of the positron with the 1s
electron of the H-like uranium ion as a function of the emitted photon
energy $\omega_1$. The unregularized cross section defined by Eq.~\eqref{DCSeq} is shown with the dashed line (blue),
whereas the solid line (red) corresponds to the regularized expression ~\eqref{dcross_IR_free}.
The top, middle, and bottom panels correspond to the kinetic positron energy of $500$ keV, $750$ keV, and $1500$ keV, respectively.
}
\label{fig_dcs}
\end{figure}

\begin{figure}[h!]
\centering
\includegraphics[width=1\linewidth]{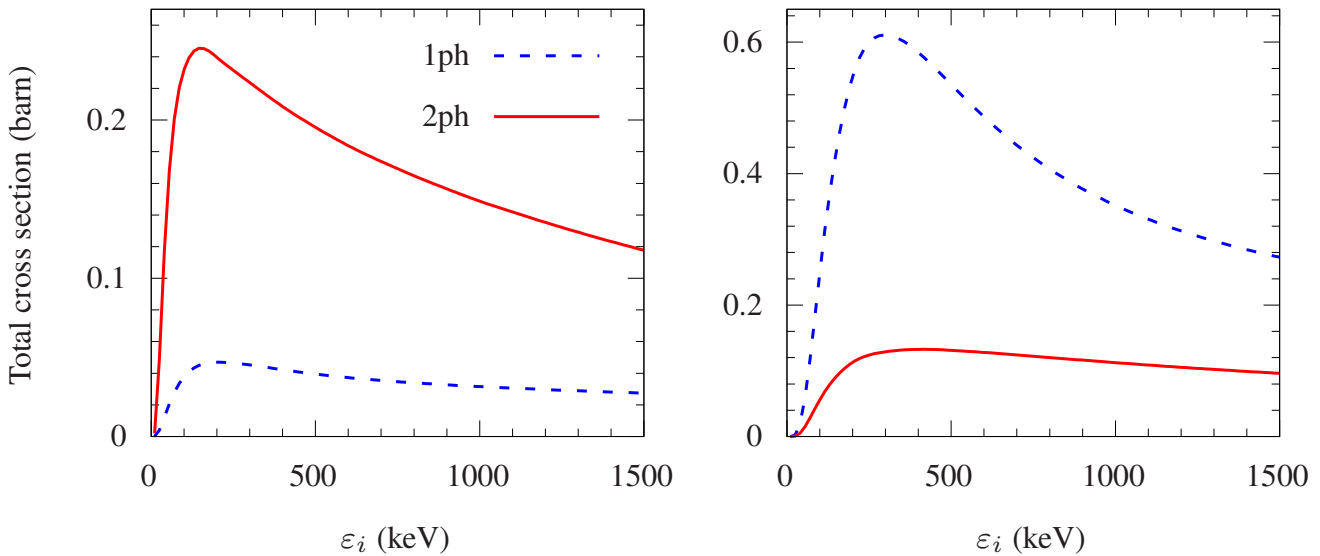}
\caption{
Total cross section for the one- and two-photon annihilation of the positron with the $1s$ electron of the H-like xenon (left panel) and uranium (right panel) ions.
}
\label{fig_Z5492}
\end{figure}

\begin{figure}[h!]
\centering
\includegraphics[width=0.98\linewidth]{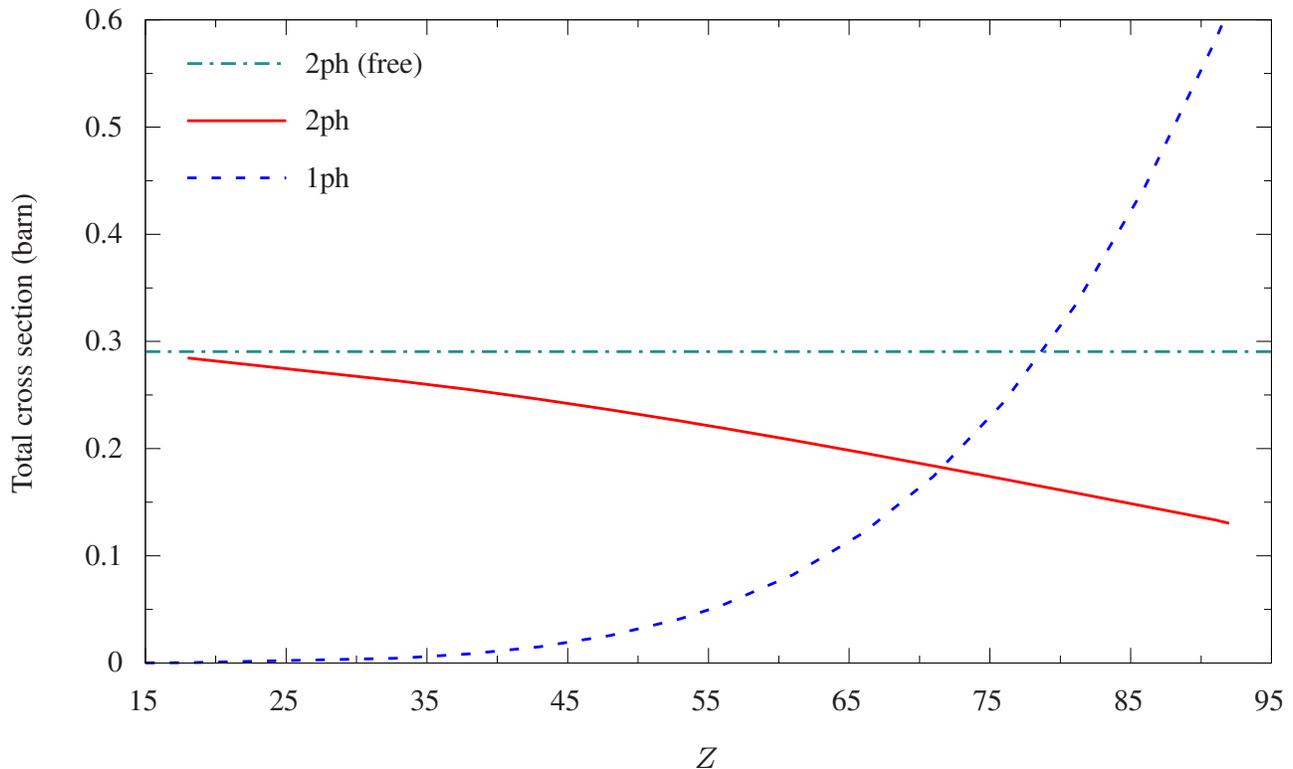}
\caption{
Total cross section for the one- and two-photon annihilation of $300$ keV positron with the $1s$ electron of the H-like ion. 
The TCS for the positron-electron annihilation in an empty space, which is given by Eq.~\eqref{eq_2ph_free}, is represented by the green dash-dotted line.
}
\label{fig_Z12}
\end{figure}

We start with presenting results for the unregularized differential cross section (DCS) for the two-photon annihilation~\eqref{DCSeq}, which contains the IR divergences.
The dependence of the DCS for the annihilation of $500$ keV, $750$ keV, and $1500$ keV positrons with the $1s$ electron of the H-like uranium ion ($Z = 92$) as a function of the emitted photon energy $\omega_1$ is presented by dashed line in Fig.~\ref{fig_dcs}.
From the figure, one can see that the DCS is symmetric with respect to the interchange of the photon energies $\omega_1 \leftrightarrow \omega_2 = E_{\rm tot} - \omega_1$.
This fact is explained by the indistinguishability of the emitted photons.
Figure~\ref{fig_dcs} clearly displays the IR divergences at the endpoints where one of the emitted photons is soft.
\\ 
\indent
To obtain meaningful results for the total cross section, one needs to eliminate
the IR divergences. This can be achieved by using the regularized cross section
given by Eq.~\eqref{ir_contrib}. We find it instructive to cross-check the analytical formula
for the divergent contribution by a numerical calculation of the unregularized
expression ~\eqref{DCSeq}. 
For this purpose, we calculate numerically the value 
\begin{equation}
\label{intensity_num}
I_{p_i}^{(\rm num)}(\omega_1) =
\frac{\omega_1}{\sigma^{(\rm 1ph)}_{\rm tot}}
\frac{d\sigma^{\rm (2ph)}}{d\omega_1},
\end{equation}
and compare its limit at $\omega_1 \rightarrow 0$ with the value provided by the analytical expression~\eqref{intensity}.
Due to the complexity of the numerical calculations at small photon energies, we restrict ourselves to the case of 500 keV positrons annihilating with the $1s$ electrons of the H-like uranium ions.
Additionally, we do not perform the calculations for $\omega_1 < 10^{-4}E_{\rm tot}$, where numerical instabilities do not allow to obtain reliable results.
Table~\ref{tab_comp} presents $I_{p_i}^{(\rm num)}(\omega_1)$ in the velocity and length gauges.
\begin{table}[h!]
\centering
\caption{
$I_{p_i}^{(\rm num)}(\omega_1)$, defined by Eq.~\ref{intensity_num}, in the velocity (second column) and length (third column) gauges for 500 keV positrons annihilating with the $1s$ electrons of the H-like uranium ions.
The extrapolated value corresponds to the limit $\omega_1 \rightarrow 0$.
}
\begin{tabular}{
l|   
S[table-format=3.9, scientific-notation=fixed, fixed-exponent=0, round-mode=places,round-precision=9]   
S[table-format=3.9, scientific-notation=fixed, fixed-exponent=0, round-mode=places,round-precision=9]
S[table-format=3.9, scientific-notation=fixed, fixed-exponent=0, round-mode=places,round-precision=9]   
}
{$\omega_1/E_{\rm tot}$} 
        & { $I_{p_i}^{(\rm num,\ vel)}(\omega_1)$ } 
        & { $I_{p_i}^{(\rm num,\ len)}(\omega_1)$ } 
        & { $I_{p_i}$ [Eq.~\eqref{intensity}] }
\\
\hline
0.01    & 0.0023450960494
        & 0.0023450960493
        &
        \\ 
0.001   & 0.0023735295
        & 0.0023735294
        &
        \\ 
0.0005  & 0.0023754744
        & 0.0023754741
        &
        \\
0.0004  & 0.0023758677
        & 0.0023758674
        &
        \\
0.0003  & 0.0023762624
        & 0.0023762620
        &
        \\
0.0002  & 0.0023766585
        & 0.0023766578
        &
        \\
0.0001  & 0.002377056
        & 0.002377055
        &
        \\ \hline
extr    & 0.00237745(1)
        & 0.00237745(1)
        & 0.0023774545
\end{tabular}
\label{tab_comp}
\end{table}
From the table, one can see that the results obtained in the different gauges agree with each other.
Moreover, the extrapolated value of $I_{p_i}^{(\rm num)}(\omega_1)$ at $\omega_1 \rightarrow 0$ is in excellent agreement with the analytical one, which is given by Eq.~\eqref{intensity}.
In what follows, we calculate the IR divergent contributions directly through the use of Eq.~\eqref{ir_contrib} and subtract them from the DCS.
The redefined DCS [Eq.~\eqref{dcross_IR_free}], which does not contain the IR divergences, is depicted in Fig.~\ref{fig_dcs} with the solid red line.
From the figure, one can see that the divergences are eliminated at all positron energies and the total cross section for the two-photon annihilation can be directly evaluated.
\\
\indent
Now we turn to the comparison of the total cross sections for the single- and double-quanta annihilation of positrons with $1s$ electrons of H-like ions.
Firstly, we investigate the dependence of the cross sections on the positron energy for medium- and high-$Z$ ions, viz., xenon ($Z = 54$) and uranium ($Z = 92$).
%
%
The total cross section (TCS) for the one- and two-photon annihilation of positrons with $1s$ electrons of these systems is presented in Fig.~\ref{fig_Z5492} as a function of the positron energy.  
From the figure, one can see that the TCS exhibits the similar behaviour
for both processes, which can be explained as follows. 
With the growth of the energy, it is easier for the positron to overcome the nucleus repulsion and annihilate with the bound electron, thus, the cross section increases.
On the other side, the growth of the energy leads to the decrease of time when the positron and electron are close to each other, which results in the drop of the annihilation probability.
The combination of these two mechanisms explains the dependence of the cross section on the positron energy, which is observed in Fig.~\ref{fig_Z5492}, namely, the growth followed by the smooth decline.
From Fig.~\ref{fig_Z5492}, one can also see that for $Z = 54$ (left panel) the two-photon annihilation dominates over the one-photon channel at all positron energies.
\\
\indent
For $Z = 92$ (right panel) the completely opposite situation is observed, namely, the single-quantum process becomes the most probable.
Let us now study the dependence of the annihilation cross sections on $Z$.
For this purpose, in Fig.~\ref{fig_Z12} we depict the TCS for the one- and two-photon annihilation of $300$ keV positron with the $1s$ electron of the H-like ion as a function of the nuclear charge $Z$.
The energy of the positron was chosen to be $300$ keV since approximately at this energy the one- and two-photon cross sections reach their maximal values for the annihilation with the uranium ion (see the right panel in Fig.~\ref{fig_Z5492}).
In this figure, one can observe that with the decrease of $Z$ the one- and two-photon annihilation cross sections approach their empty space limits.
The single-quantum annihilation vanishes at low $Z$, which expresses the fact that the annihilation of free electron and positron with the emission of one photon is forbidden.
\\
\indent
The TCS for the two-quantum annihilation turns into the cross section for the analogous channel in an empty space, which is given by~\cite{Akhiezer_1965}:
\begin{equation}
\label{eq_2ph_free}
\sigma^{(\rm 2 ph,\ free)}_{\rm tot} 
= 
\pi \alpha^2 \frac{1}{\varepsilon_i + 1}\left[
\frac{\varepsilon_i^2+4\varepsilon_i+1}{\varepsilon_i^2 - 1}\ln{\left(\varepsilon_i+\sqrt{\varepsilon_i^2 - 1}\right)} - \frac{\varepsilon_i+3}{\sqrt{\varepsilon_i^2 - 1}}
\right]
\end{equation}
and depicted in Fig.~\ref{fig_Z12} by the green dash-dotted line.
From Fig.~\ref{fig_Z12}, it is also seen that with the growth of $Z$ the one-photon annihilation increases while the two-photon one decreases.
For a wide range of $Z$, the double-quanta channel prevails over the single-quantum one, but for heavy systems with $Z > 70$, the situation becomes the opposite.

%% file: sections/summary.tex
\section{Conclusion}
\label{sec5}
We have studied the process of the two-photon annihilation of a positron with an electron bound in the $1s$ state of an H-like ion.
The calculation was performed within the fully relativistic QED formalism, with the nuclear binding field accounted for in a nonperturbative manner.
The complete spectrum of intermediate electron-positron Dirac states in the binding field of the nucleus was described by the exact Dirac-Coulomb Green function.
The IR-divergent contributions, which occur in the situations when one of the emitted photons is soft, were calculated separately and subtracted from the differential cross section for the two-photon annihilation.
\\
\indent
The developed approach was applied to the calculation of the total cross section for the double-quanta annihilation of positrons with the $1s$ electrons of H-like ions in a wide range of the positron energy and the nuclear charge number $Z$.
The cross sections of the one-photon and two-photon annihilation channels were compared for different values of $Z$.
We have demonstrated that for the low- and medium-$Z$ ions the two-photon annihilation dominates over the one-photon channel for all positron energies.
The situation becomes reversed for heavy ions, such as uranium ($Z = 92$).
The probabilities of the single- and double-quanta channels are shown to be approximately equal to each other in the region $Z \approx 70$ for the 300~keV positrons.
\\
\indent
The formalism developed in the present work substantially extends the domain of collision energies available for an accurate theoretical description of the two-photon annihilation. Detailed theoretical and experimental investigations of this process will
help to unfold various features of the annihilation phenomenon in the presence of the strong Coulomb field. 

%

%% file: sections/acknowledgements.tex
\section*{Acknowledgements} 
\label{sec:acknowledgements}
This study was supported by the grant of the Russian Science Foundation \textnumero 22-22-00370,
\url{https://rscf.ru/project/22-22-00370/}.